\documentclass[aps,twocolumn]{revtex4}

\usepackage{graphicx}
\usepackage{graphics}
\usepackage{epstopdf}
\def\bsigma{\mbox{\boldmath $\sigma$}}
\def\bnabla{\mbox{\boldmath $\nabla$}}

\begin{document}
\title{ Flux noise in a superconducting transmission line}
\author{F. T. Vasko  }
\email{ftvasko@gmail.com}
\affiliation{QK Applications, San Francisco, CA 94033, USA }
\date{\today}

\begin{abstract}
We study a superconducting transmission line (TL) formed by distributed LC oscillators and excited by external magnetic fluxes which are aroused from random magnetization (A) placed in substrate or (B) distributed at interfaces of a two-wire TL. Low-frequency dynamics of a random magnetic field is described based on the diffusion Langevin equation with a short-range source caused by (a) random amplitude or (b) gradient of magnetization. For a TL modeled as a two-port network with open and shorted ends, the effective magnetic flux at the open end has non-local dependency on noise distribution along the TL. The flux-flux correlation function is evaluated and analyzed for the regimes (Aa), (Ab). (Ba), and (Bb). Essential frequency dispersion takes place around the inverse diffusion time of random flux along the TL. Typically, noise effect increases with size faster than the area of TL. The flux-flux correlator can be verified both via the population relaxation rate of the qubit, which is formed by the Josephson junction shunted by the TL with flux noises, and via random voltage at the open end of the TL.  
\end{abstract}
\maketitle

\section{Introduction}
Recent progress in the implementation and the benchmarking of the quantum-information-processing protocols is based on different types of superconducting flux qubits connected through TLs, which provide inter-qubit  links, control of qubit's states, and readout, see \cite{1,2,3,4} and the references therein. These elements form an essential part of quantum hardware and their effect on the fidelity of different computational protocols should be elucidated. Although the dynamic properties of multi-qubit devices have been analyzed with the use of the lumped-element approach, \cite{5} the effects of  environment on quantum hardware has not been investigated completely. Superconducting circuits have been studied extensively for sensor applications at temperatures around 1 K, \cite{6} where noise arises due to two-level defects. For operating temperatures of quantum hardware, near $10^{-2}$ K, microscopic mechanisms of interaction with flux noises are under investigation now, see a detailed discussion \cite{3} and recent experimental data \cite{7,8} with references therein. In particular, mechanisms of low-frequency flux noise in TL have not been analyzed and their effects on quantum hardware, including spectral and size dependencies, have not been characterized. Thus, it is timely now to study flux noises caused by random magnetization around TLs and to consider physical effects caused by these noises.
\begin{figure}
\begin{center}
\includegraphics[scale=0.45]{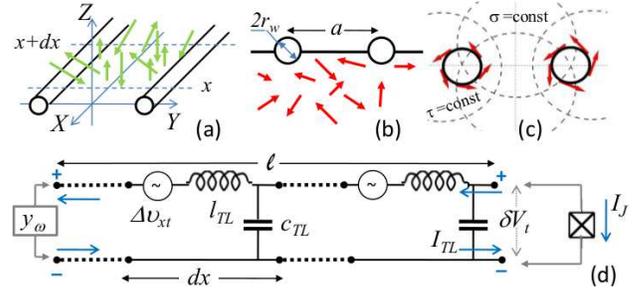}
\end{center}
\addvspace{-0.5 cm}
\caption{(a) Geometry of a two-wire TL with random external flux (the green arrows) penetrated through the interval ($x,x+dx$). (b) TL placed on substrate with random magnetization (the red arrows); $a$ is the distance between the wires of radius $r_w$. (c) Y0Z cross section of a two-wire TL with fluctuations of tangential magnetization along the wires (the red arrows). Gray curves sketch the bipolar coordinate system ($\tau ,\sigma$). (d) Circuit model for a TL of length $\ell$ under low-frequency noise. Part of the TL between $x$ and $x+dx$ with inductance $l_{TL}dx$, capacitance $c_{TL}dx$, and random drop of potential $\Delta v_{xt}$ is shown in the left cell. Shunting of the TL via two-pole scheme of admittance $y_\omega$ and via a Josephson junction is shown at the left and the right, respectively. }
\end{figure}

In this paper, we consider a two-wire TL (Fig. 1a) formed by distributed LC oscillators under the influence of different types of low-frequency flux noises. Based on the standard electrodynamic approach, \cite{9,10} we consider the TL as a two-port network and evaluate a random e.m.f. due to fluctuations of magnetic flux across the area of the TL. Within the low-frequency approximation, when the characteristic frequency of TL $\omega_{LC}$ is greater than $\omega$, this evaluation gives {\it an  effective non-local flux} through a TL of length $\ell$ as $\Delta \Phi_{xt} |_{x=-\ell /2}^{x=\ell /2}+\ell (d\Delta \Phi_{xt}/dx)\left |_{x=-\ell /2}^{x=\ell /2}\right.$ with random flux along the TL, $\Delta \Phi_{xt}$, and its derivative taken at the ends of TL, $x=\pm\ell /2$, see Eq. (\ref{eq:12}) below. We consider the case of {\em classical external noises} which are aroused from random magnetization (A) placed in substrate (Fig. 1b) or (B) distributed along interfaces of the superconducting wires which form the TL (Fig. 1c). 

The low-frequency dynamics of a random magnetic field is governed by the Langevin equation with short-range sources caused by (a) random amplitude or (b) gradient of magnetization. We employ the hydrodynamic approach with an isotropic diffusion coefficient and a relaxation rate which are introduced phenomenologically. A general out-of-equilibrium regime of fluctuations is considered here, without any restrictions imposed by the fluctuation-dissipation theorem. The models of noise suggested here are consistent with recent experimental data reporting effects of random magnetization in a substrate and along interface regions, see \cite{11} and \cite{12,13} respectively, as well as with an essential stray population of qubit (hot qubits) at low temperatures, reported on in Refs. \cite{13,14,15}. Beside of this, a typical flux qubit is formed by the Josephson junction's loop shunted by long TL (not an effective LC circuit as supposed in simplified models), and a coupling of this qubit to low-frequency noise is described by the approach suggested here. Dependencies  of noise on $\omega$ and $\ell$ can be verified via dissipative dynamics of this qubit or via measurements of a random voltage appearing at the open end of the TL if the other end is shortened. These regimes of measurements are illustrated schematically in Fig. 1d and they are described by the flux-flux correlation function of noises penetrating through the TL's area. 

The obtained results demonstrate that spectral and size dependencies of the correlators are rather different for the cases (Aa), (Ab), (Ba), and (Bb). Spectral dispersion takes place if $\omega$ is comparable to the inverse diffusion time along TL, $\tau_D^{-1}$. For different cases, the correlator is decreased $\propto (\omega\tau_D )^{-1}$, $\propto (\omega\tau_D )^{-0.5}$, or $\propto\ln (1/\omega\tau_D )$ at high frequencies ($\omega\tau_D\gg 1$) and it is increased $\propto (\omega\tau_D )^{-2}$ or saturated in the low-frequency region ($\omega\tau_D\ll 1$), see below the asymptotic dependencies given by Eq. (\ref{eq:38}) and Table I in Conclusion. For fixed specific inductance and capacitance, these correlators are not proportional to the TL's area; typically, the external-noise effect increases faster than mechanisms \cite{12,13,16} caused by the RKKY interaction between near-interface spins via the superconductor and proportional to the length of the TL. Bacause of this key difference, even weak external noises may give a dominant contribution in long TLs. The analytical results obtained here open the way {\it to a verification of noise's effect on quantum hardware and a determination of phenomenological parameters} describing flux fluctuations. Characterization of noises in multi-qubit clusters, which include qubits, the inter-qubit connections, and the qubit's control lines,  opens the way to improving the fidelity of the hardware for quantum-information  processing. 

The rest of the paper is organized as follows. The effects of low-frequency noise on TLs are described in Sec. II. The flux-flux correlation functions for the mechanisms of noise under consideration  are analyzed in Sec. III. Concluding remarks with a summary of spectral and size dependencies, a list of assumptions, and an outlook are given in the last section. The dissipative dynamics of a flux qubit shunted by a TL and the integrals used in Sect. IIIB are considered in Appendixes A and B, respectively. 

\section{Effect of noise on a shunted TL }  
Based on the lumped-element approach, we analyze here the effective circuit for TL under random magnetic flux arising from the above-listed mechanisms. For a TL of length $\ell$ ($|x|<\ell /2$), distributions of voltage and current along the TL, $V_{xt}$ and $I_{xt}$, are connected by the system of equations \cite{5,9} 
\begin{eqnarray}
\frac{{\partial V_{xt} }}{\partial x} + l_{TL}\frac{{\partial I_{xt} }}{{\partial t}} =
\frac {\partial\Delta v_{xt}}{\partial x} ~,  \nonumber  \\
\frac{{\partial I_{xt} }}{\partial x} + c_{TL}\frac{{\partial V_{xt} }}{{\partial t}} = 0 ~.
\label{eq:1}
\end{eqnarray} 
Here $l_{TL}=L/\ell$ and $c_{TL}=C/\ell$ are the inductance and the capacitance per unit length, which are written through the total inductance and capacitance, $L$ and $C$. Performing an integration of the Maxwell equation ${\rm curl}{\bf E}=\ldots$ over inter-wire area $S=a\times dx$ (see Fig. 1a) we obtain the random e.m.f. in Eq. (\ref{eq:1}), $\partial\Delta v_{xt}/\partial x$, as follows
\begin{equation}
\left( \frac{\partial\Delta v_{xt}}{\partial x}\right) dx=-\frac{1}{c}\frac{\partial}{\partial t}\int_{(S)}d{\bf S}\cdot{\bf h}_{{\bf r}t} ~,  \label{eq:2}
\end{equation}
where ${\bf h}_{{\bf r}t}$ is the external magnetic field penetrating through two-wire TL; see the standard electrodynamic derivation  of the telegraph equations. \cite{9,10} The boundary conditions at $x=\pm\ell /2$ determine a connection between the edge voltages and currents, $V_{x=\pm\ell /2,t}\equiv V_{\pm ,t}$ and $I_{x=\pm\ell /2,t}\equiv I_{\pm ,t}$. After the Fourier transform in the $t$-domain and substitution of $I_{x\omega}$ through $dV_{x\omega}/dx$ in Eqs. (\ref{eq:1}) one obtains the second-order telegraph equation 
\begin{equation}
\left( {\frac{d^2}{dx^2} + k_\omega ^2 } \right)V_{x\omega }  = \frac{d^2 \Delta v_{x\omega}}{dx^2}\equiv w_{x\omega }  ,  \label{eq:3}
\end{equation}
where $k_\omega =\omega\sqrt{l_{TL}c_{TL}}=\omega /(\omega_{LC}\ell )$ is the wave vector at frequency $\omega$ for TL with the characteristic frequency $\omega_{LC}=1/\sqrt{LC}$. 

From the upper line of Eq. (\ref{eq:1}) we obtain $I_{x\omega}$ through voltages $V_{x\omega}$ and $\Delta v_{x\omega}$ as
\begin{equation}
I_{x\omega}=\frac{i}{\omega l_{TL}}\left(\frac{dV_{x\omega}}{dx}-\frac{d\Delta v_{x\omega}}{dx}\right) ~.  \label{eq:4}
\end{equation}
The voltage and current distributions along the TL are given by
\begin{eqnarray}
V_{x\omega }=v_s \sin k_\omega x+v_c\cos k_\omega x-\Delta V_{x\omega } ~, ~~~   \\
I_{x\omega }  = \frac{{ik_\omega  }}{{\omega l_{TL} }}\left( {v_s \cos k_\omega  x - v_c \sin k_\omega  x} \right) - \Delta I_{x\omega } ~,  \nonumber   \label{eq:5}
\end{eqnarray}
where we separate the noise-induced contributions, $\Delta V_{x\omega}$ and $\Delta I_{x\omega}$, which are written in the form 
\begin{eqnarray} 
\Delta V_{x\omega}=\sin k_\omega x\int_x^{\ell /2}\frac{dx'}{k_\omega}\cos k_\omega x'w_{x'\omega}  \nonumber   \\ + \cos k_\omega x\int_{-\ell /2}^x \frac{dx'}{k_\omega} \sin k_\omega  x'w_{x'\omega} ~, ~~~~  \label{eq:6} \\ 
\Delta I_{x\omega }  = \frac{i}{\omega l_{TL} }\left( \cos k_\omega  x\int_x^{\ell /2} dx'\cos k_\omega  x'w_{x'\omega }  \right. \nonumber  \\ 
\left. - \sin k_\omega  x\int_{-\ell /2}^x dx'\sin k_\omega  x'w_{x'\omega }  + \frac{d\Delta v_{x\omega}}{dx} \right) ~. \nonumber  
\end{eqnarray} 
The constants $v_{s,c}$ in Eqs. (5) are determined through edge voltages $V_{\pm ,\omega}$ and random contributions $\Delta V_{\pm\ell /2,\omega}$ according to
\begin{equation}
 \left| {\begin{array}{*{20}c} {v_s }  \\ {v_c } \end{array}} \right| =(\sin k_\omega\ell )^{-1} \left| {\begin{array}{*{20}c}
   {\cos (k_\omega\ell /2){\cal V}_-}  \\
   {\sin (k_\omega\ell /2){\cal V}_+} \end{array}} \right| ~,  \label{eq:7}
\end{equation}
where we introduce ${\cal V}_{\pm}\equiv V_{+,\omega}+\Delta V_{\ell /2,\omega}\pm V_{-,\omega}\pm\Delta V_{-\ell /2,\omega}$. After substitution of $v_{s,c}$ into Eqs. (\ref{eq:5}) one connects the edge currents $I_{\pm ,\omega}$ and voltages $V_{\pm ,\omega}$:
\begin{eqnarray} 
\left| \begin{array}{*{20}c}
I_{+,\omega }  \\   I_{-,\omega}  \\
\end{array} \right| = \hat Y_\omega  \left| \begin{array}{*{20}c}
V_{+,\omega }+\Delta V_{\ell /2,\omega }  \\
   V_{-,\omega }  + \Delta V_{-\ell /2,\omega}   \\
\end{array}\right| - \left| \begin{array}{*{20}c}
  \Delta I_{\ell /2,\omega }  \\
   \Delta I_{-\ell /2,\omega }  \\
\end{array} \right| ~,  \nonumber  \\
\hat Y_\omega =\frac{ik_\omega}{\omega l_{TL}\sin k_\omega\ell}\left| {\begin{array}{*{20}c} \cos k_\omega\ell & { - 1}  \\ 1 & - \cos k_\omega\ell \end{array}} \right| ~. ~~~~ \label{eq:8}
\end{eqnarray}
Here $\hat Y_\omega$ is the admittance matrix \cite{17} and the noise contributions, $\Delta V_{\pm\ell /2,\omega}$ and $\Delta I_{\pm\ell /2,\omega}$, are given by integrals over the TL determined by Eqs. (\ref{eq:6}).

Below we consider a TL which is shunted by a lossless two-pole circuit with the imaginary admittance $y_\omega$ connected at $x=-\ell /2$, when $I_{-,\omega }=y_\omega V_{-,\omega}$. Using this relation and Eqs. (\ref{eq:8}) we write  the current-voltage relation at the other end of the TL, $x=\ell /2$, as
\begin{equation}
I_{+,\omega}={\cal Y}_\omega V_{+,\omega}-\delta{\cal J}_\omega , ~~~ {\cal Y}_\omega  = Y_{11}+\frac{Y_{12} Y_{21}}{y_\omega -Y_{22}} ~.  \label{eq:9}
\end{equation}
Here we introduce the effective admittance of shunted TL, ${\cal Y}_\omega$, and the noise-induced random current $\delta{\cal J}_\omega$ at $\ell /2$ is written through the solutions (\ref{eq:6}) at $x=\pm\ell /2$ as follows
\begin{eqnarray} 
\delta{\cal J}_\omega = \Delta I_{\ell /2,\omega} -Y_{11}\Delta V_{\ell /2,\omega}- Y_{12}\Delta V_{-\ell /2,\omega}  ~~~~  \label{eq:10} \\
-\frac{Y_{12}}{y_\omega -Y_{22}}\left(\Delta I_{-\ell /2,\omega} -Y_{21}\Delta V_{\ell /2,\omega}- Y_{22}\Delta V_{-\ell /2,\omega}\right)   ~.  \nonumber  
\end{eqnarray}
For the case of the open edge condition at $x=\ell /2$, when $I_{+,\omega}=0$, Eq. (\ref{eq:9}) gives the random voltage as $\delta V_\omega =\delta{\cal J}_\omega /{\cal Y}_\omega$ and the voltage-voltage correlation function takes the form:
\begin{equation}
\left\langle {\delta V\delta V} \right\rangle_\omega =\left\langle {\delta{\cal J}\delta{\cal J}} \right\rangle_\omega /|{\cal Y}_\omega |^2 ~,  \label{eq:11}
\end{equation}
where $\left\langle\ldots\right\rangle$ means averaging over a random e.m.f. The spectral dependencies of $\left\langle {\delta V\delta V} \right\rangle_\omega$ are determined by both the correlations of the random contributions at $\pm\ell /2$ given by Eq. (\ref{eq:6})  and the admittances $\hat Y_\omega$ and $y_\omega$. Thus, a random e.m.f. along a TL shunted at $-\ell /2$ gives rise to a random current $\delta{\cal J}_\omega$ at $\ell /2$ which leads to the voltage fluctuations described by Eq. (\ref{eq:11}) and modifies the dissipative dynamics of qubit considered in Appendix A. Below, we restrict ourselves to the case of a TL shortened at $x=-\ell /2$, where the edge condition is $V_{-,\omega}\to 0$ because $|y_\omega |\to\infty$, so ${\cal Y}_\omega\simeq Y_{11}$ and $\delta{\cal J}_\omega$ is determined by the upper line of Eq. (\ref{eq:10}).

In the low-frequency region, $k_\omega\ell =|\omega |/\omega_{LC}\ll 1$, we use the series expansion $\cot z\approx 1/z-z/3-z^3 /45-\ldots$ in the diagonal elements of $\hat Y_\omega$ given by Eq. (\ref{eq:8}) and the effective admittance determined by Eq. (\ref{eq:9}) takes the form ${\cal Y}_\omega\approx i\left( 1/\omega L -\omega C/3\right)$.  Introducing the random flux $\Delta\Phi_{x\omega}=\Delta v_{x\omega }/i\omega$ and performing integrations by parts in $\Delta V_{\ell /2,\omega}$ and $\Delta I_{\pm\ell /2,\omega}$ according to Eqs. (\ref{eq:6}), one transforms Eq. (\ref{eq:10}) into 
\begin{equation} 
\delta\Phi_t\! \equiv\! -L\delta{\cal J}_t\!\approx\! \Delta \Phi_{x\omega } |_{x=-\ell /2}^{x=\ell /2}+\ell \frac{d\Delta \Phi_{x\omega}}dx\left|_{x=-\ell /2}^{x=\ell /2}\right.   ,  \label{eq:12}
\end{equation}
where $\delta\Phi_t$ is the effective flux at $x=\ell /2$ and the $\propto (\omega /\omega_{LC})^2$ contributions are omitted. As a result, the voltage-voltage correlator (\ref{eq:11}) at the open end of the TL is written through the flux-flux correlator, $\left\langle {\delta V\delta V} \right\rangle_\omega =\omega^2 \left\langle {\delta\Phi\delta\Phi} \right\rangle_\omega$, so the fluctuations of voltage are suppressed if $\omega\to 0$ for cases (Aa), (Ab), and (Bb), see below.

\section{Flux-flux correlation functions } 
Random voltage at the open end of the TL and dissipative dynamics of a qubit shunted by the TL are  determined by the correlation function $\left\langle {\delta\Phi_{t + \Delta t} \delta\Phi_t} \right\rangle$. In order to calculate this correlator using Eqs. (\ref{eq:2}) and (\ref{eq:12}), we consider the Fourier transform of the flux distribution $\Delta \Phi_{xt}$, which is determined from the relation
\begin{equation}
\left(\frac{d\Delta\Phi_{x\omega}}{dx}\right) dx=-\frac{1}{c}\int_{(S)} d{\bf S}\cdot{\bf h}_{{\bf r}\omega} ~.  \label{eq:13}
\end{equation}
Next, we describe the transverse to the TL component of a low-frequency magnetic field, $h_{{\bf r}t}^\bot$, with the Langevin equation by taking into account diffusion and magnetization damping. Within the isotropic approximation, the  Langevin equation is written using the scalar diffusion coefficient and relaxation rate, $D$ and $\nu$, in the form
\begin{equation}
\left(\frac{\partial }{\partial t}-D\nabla_r^2 
+\nu \right)h_{{\bf r}t}^\bot =\zeta_{{\bf r}t}^\bot ~.  \label{eq:14}
\end{equation}
Random source $\zeta_{{\bf r}t}^\bot$ is determined by the short-range correlation function in the spatiotemporal domain,
\begin{equation}
\left\langle {\zeta _{{\bf r}t}^ \bot\zeta _{{\bf r}'t'}^\bot}\right\rangle =\left[ W+ \overline W(\bnabla_r\cdot \bnabla_{r'})\right] \delta({\bf r} - {\bf r}')\delta (t-t') ~,
  \label{eq:15}
\end{equation}
where $W$ and $\overline W$ determine level of fluctuations caused by random amplitude and gradient of magnetization, respectively. \cite{18}  Below, we analyze the models of noise in lower half-space (A) or at interfaces (B), when $\bnabla_r$ is 3D- or 2D-gradient, and we restrict ourselves to the case of a long and wide TL, $\ell\gg a\gg r_w$. We consider the region $\omega\geq\nu$ because $ \left\langle {\delta \Phi \delta \Phi}\right\rangle_\omega$ is $\omega$-independent if $\omega\ll\nu$. 

\subsection{Noise excited in a substrate}
For the case of substrate with a random magnetic field shown in Fig. 1b, Eqs. (\ref{eq:12}) and (\ref{eq:13}) suggest that effective flux $\delta \Phi _t$ is obtained through the transverse component of field $h_{{\bf r}t}^\bot=({\bf h}_{{\bf r}t}\cdot{\bf e}_z )$ at interface $z=0$ according to
\begin{equation}
\delta \Phi _t  =  -\int\limits_{ - a/2}^{a/2} \frac{dy}{c} \left( {\left.\ell{h_{{\bf r}t}^ \bot  } \right|_{x=-\ell /2}^{x =\ell /2}  + \int\limits_{-\ell /2}^{\ell /2} {dx} h_{{\bf r}t}^ \bot  } \right)_{z=0} .  \label{eq:16}
\end{equation}
The random field $h_{{\bf r}t}^\bot$ is governed by Eq. (\ref{eq:14}) with the zero-flow boundary condition $(\partial h_{{\bf r}t}^\bot/\partial z)_{z=0}=0$. The symmetry with respect to the $z=0$ solution of Eq. (\ref{eq:14}) is written with the bulk Green's function $G_{\Delta{\bf r}\Delta t}$ and a spatiotemporal correlator of random fields is given by
\begin{eqnarray}
\left\langle h_{{\bf r}t}^ \bot  h_{{\bf r}'t'}^ \bot  \right\rangle_{z,z'=0}=\int\!\!\int_{(z_1,z'_1 < 0)}\!\!\!\! d{\bf r}_1 d{\bf r'}_1\int\!\!\int\!\! dt_1 dt'_1 \nonumber   \\
\times G_{{\bf r}-{\bf r}_1 t -t_1} G_{{\bf r}' - {\bf r'}_1t'-t'_1}\left\langle {\zeta_{{\bf r}_1 t_1}^ \bot\zeta _{{\bf r'}_1 t'_1}^\bot}\right\rangle ~,~~~~~ \label{eq:17}
\end{eqnarray} 
where ${\bf r}=({\bf r}_\| ,z)$. Using Eq. (\ref{eq:15}), performing integration by parts over ${\bf r}_1 ,{\bf r'}_1$, and introducing $\Delta t=t-t'$ one obtains
\begin{eqnarray}
\left\langle h_{{\bf r}t}^ \bot  h_{{\bf r}'t'}^ \bot \right\rangle_{z,z'=0}\!\! =\!\!\!\! \!\!\int\limits_{(z_1 < 0)}\!\!\!\!\!\! d{\bf r}_1\!\!\int\!\! dt_1\!\!\left[ WG_{{\bf r}-{\bf r}_1\frac{\Delta t}{2}-t_1}G_{{\bf r}'-{\bf r}_1 ,-\frac{\Delta t}{2}-t_1} \right. \nonumber   \\
\left. +\overline W (\bnabla_{r_1}G_{{\bf r}-{\bf r}_1 ,\frac{\Delta t}{2}-t_1})\cdot (\bnabla_{r_1}G_{{\bf r}'-{\bf r}_1,-\frac{\Delta t}{2}-t_1}\right] . ~~~~~~ \label{eq:18}
\end{eqnarray} 
After substitution of the Fourier transform of the Green's function $G_{{\bf k}t}=\theta (t)\exp (-Dk^2t)/(2\pi )^2$ and integrations over time $t_1$ and in-plane coordinates $({\bf r}_1 )_{\|}$, we re-write this correlator in the form [here $\Delta{\bf r}_{\|}={\bf r}_{\|}-{\bf r}'_{\|}$ and ${\bf k}=({\bf k}_\| ,k_\bot )$] 
\begin{eqnarray}
\left\langle h_{{\bf r}t}^ \bot  h_{{\bf r}'t'}^ \bot \right\rangle_{z,z'=0}\!\! =\frac{{e^{ - \nu |\Delta t|} }}{D}\!\!\int\limits_{-\infty }^0\!\!{dz}\!\!\int\!\! {dk_ \bot  } \!\!\int\!\! {dk'_ \bot} e^{i(k_ \bot   + k'_ \bot  )z} ~~~~   \\
\times\!\! \int\!\!\frac{d{\bf k}_{\|} }{(2\pi )^3 } \frac{e^{i{\bf k}_{\|}\cdot\Delta {\bf r}_{\|} - D(k_{||}^2  + k_\bot ^2 )|\Delta t|}}{2k_{\|}^2  + k_\bot ^2 + k'_\bot{}^2 }\left[ {W + \overline W(k_{\|}^2 -k_\bot k'_\bot )} \right] .  \nonumber  \label{eq:19}
\end{eqnarray} 
Furthermore, we integrate over $k'_\bot$ and the transverse coordinate $z$, so the  correlator of the random fields is transformed into
\begin{eqnarray}
 \left\langle h_{{\bf r}t}^\bot  h_{{\bf r}'t'}^\bot \right\rangle_{z,z'=0}\!\! 
 = \frac{\pi e^{-\nu |\Delta t|}}{D}\!\!\int\!\! \frac{d{\bf k}}{(2\pi )^3} \frac{W +\overline Wk^2}{2 k^2}  \nonumber  \\ 
\times\exp (i{\bf k}_{\|}\cdot\Delta{\bf r}_{\|}-Dk^2 |\Delta t|) ~. \label{eq:20}
\end{eqnarray}

The flux-flux correlator $\left\langle {\delta\Phi\delta\Phi} \right\rangle_\omega$
in Eqs. (\ref{eq:11}) and (\ref{eq:A6}) is obtained after the Fourier transform over $\Delta t$, which gives $2(Dk^2 +\nu) /[(Dk^2 +\nu )^2 +\omega^2 ]$, and integrations over the X0Y-plane, according to Eq. (\ref{eq:16}). The integration over the area of the TL gives the form-factor $\Psi\left( X,Y\right) = (\sin X\sin Y)^2 [1+(2X)^2 ]/(XY)^2$ and the correlator takes the final form:
\begin{eqnarray}
\left\langle {\delta\Phi\delta\Phi} \right\rangle_\omega = \frac{{\pi (a\ell )^2 }}{{Dc^2 }}\int {\frac{{d{\bf k}}}{{(2\pi )^3 }}} \Psi \left( {\frac{k_x\ell }{2},\frac{k_y a}{2}} \right)
\nonumber  \\
\times\frac{{(Dk^2  + \nu )(W +\overline Wk^2 )}}{{k^2 \left[ {(Dk^2  + \nu )^2  + \omega ^2 } \right]}} ~. ~~~~~~   \label{eq:21}
\end{eqnarray}
Using the dimensionless wave vector ${\bf u}={\bf k}\sqrt{a\ell}/2$ and averaging over the angles of the form-factor, $\psi_u =\int d\Omega_{\bf u}\Psi (u_x ,u_y /p)/4\pi$ which is dependent on the aspect ratio $p=\ell /a$, one arrives at the integral over $u$:  
\begin{equation}
 \left\langle {\delta\Phi \delta\Phi}\right\rangle_\omega = \int\limits_0^{u_{\max }} du \psi_u \frac{(u^2 +\nu\tau_D )(A+B u^2 )}{(u^2  + \nu \tau_D )^2  + (\omega \tau_D )^2}  ~.  
\label{eq:22}
\end{equation}
Here, we introduce the diffusion time along the TL $\tau_D =\ell^2 /4D$, the coefficients $A=W\ell (a\ell /Dc)^2$ and $B=\overline W\ell (2a/Dc)^2$, which determine strength of noises, and a cut-off factor $u_{\max }$ which is due to the inapplicability of Eq. (\ref{eq:14}) for short scales.
\begin{figure}
\begin{center}
\includegraphics[scale=0.21]{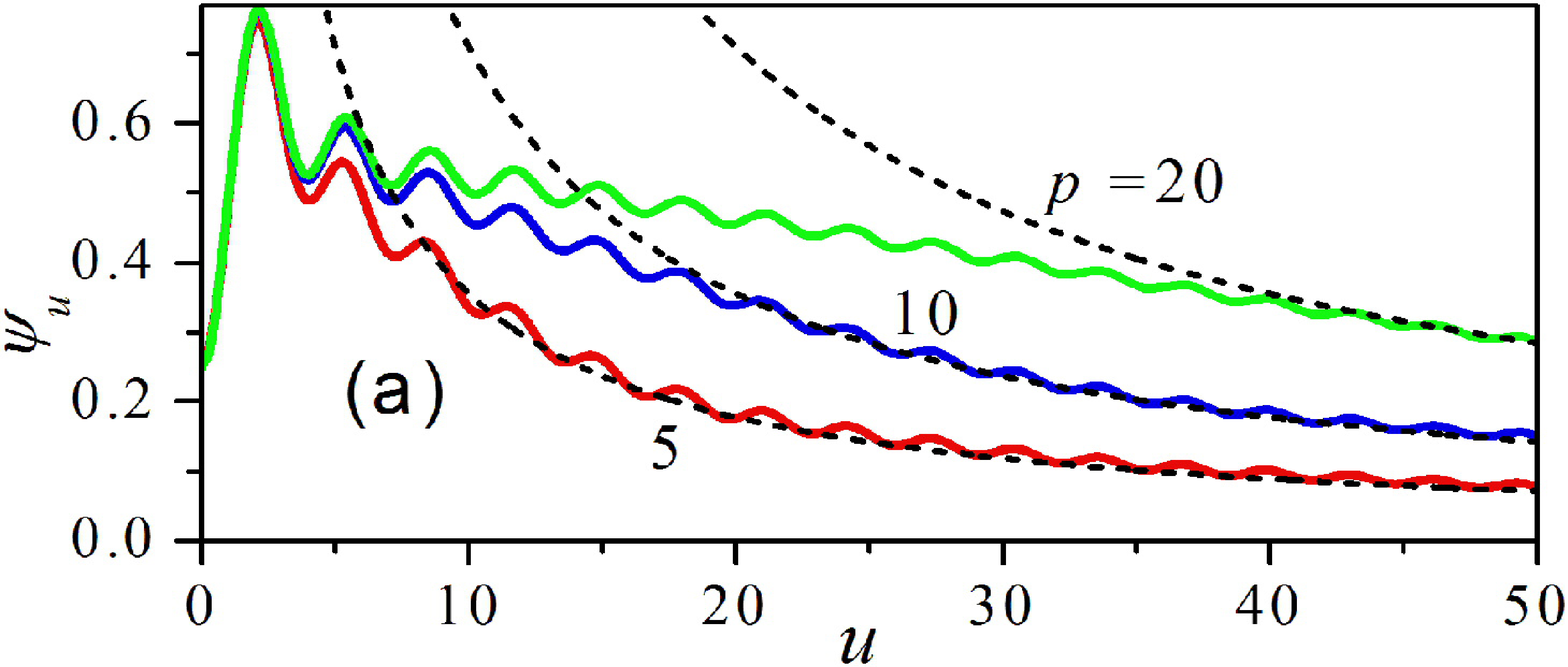}
\includegraphics[scale=0.21]{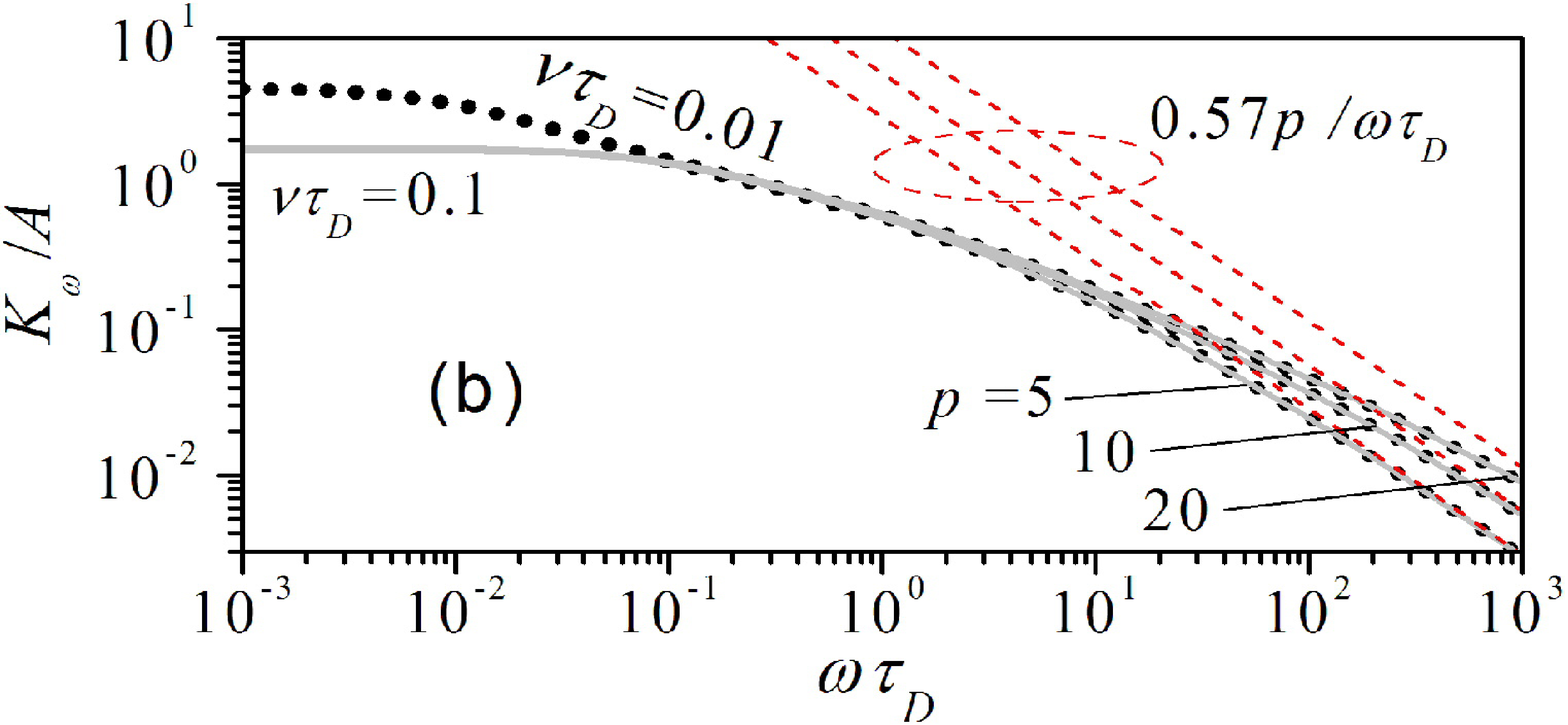}
\includegraphics[scale=0.21]{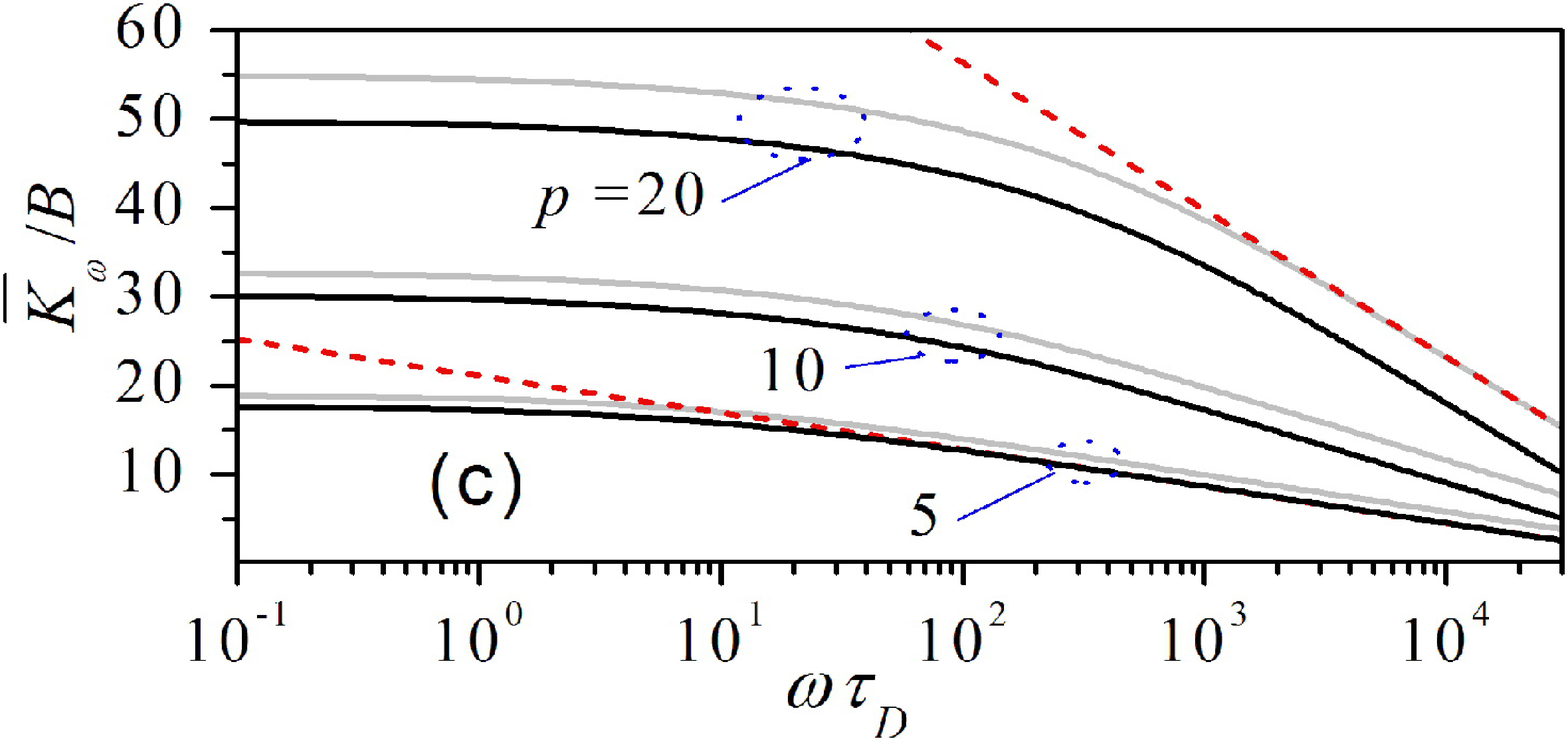}
\end{center}
\addvspace{-0.5 cm}
\caption{ (a) Averaged form-factor $\psi_u$ versus the dimensionless wave vector $u=k\ell /2$ for aspect ratios $p=$5 (red), 10 (blue), and 20 (green). Dashed curves show asymptotic behaviors determined by $0.71 p/u$. 
(b) Flux-flux correlation function $K_\omega /A$ versus the dimensionless frequency $\omega\tau_D$ for the case (Aa) with $\nu\tau_D$=0.01 (the black dotted curves) and 0.1 (the gray curves). The red dashed lines show asymptotes for $\omega\tau_D\gg 1$. (c) Function $\overline K_\omega /B$ [case (Ab)] versus $\omega\tau_D$ for $\nu\tau_D$=0.05, $p=$5, 10, or 20 (marked), and $u_{\rm max}=$350 (black) or 500 (gray) plotted in semi-log scale. The red logarithmic asymptotes correspond $(u_{\rm max},p)=$(350,5) and (500,20). }
\end{figure} 

The averaged form-factor $\psi_u$ is shown in Fig. 2a together with the hyperbolic asymptotes $\psi_u \approx 0.72 p/u$ which are valid at $u>\widetilde u\sim 2p$. Separating contributions (a) and (b) as the sum $\left\langle {\delta\Phi\delta\Phi}\right\rangle_\omega =K_\omega +\overline K_\omega$ and integrating over the tail region $(\widetilde u,u_{\max })$, we transform Eq. (\ref{eq:22}) into
\begin{eqnarray} 
\left| {\begin{array}{*{20}c} K_\omega /A  \\ \overline K_\omega /B \end{array}} \right| = \int\limits_0^{\widetilde u} {du} \frac{\psi _u (u^2 +\nu \tau_D )}{{(u^2  + \nu \tau _D )^2  + (\omega \tau _D )^2 }}\left| {\begin{array}{*{20}c} 1  \\ {u^2 } \end{array}} \right|  \nonumber  \\
+0.36p \left| {\begin{array}{*{20}c}\left[\pi /2-\arctan\left( {\widetilde u}^2 /{\omega\tau_D}\right)\right] /\omega\tau_D \\  \ln (u_{\rm max}^2 )-\ln{\sqrt{ (\omega\tau_D )^2+\widetilde u^4}}  \end{array}} \right|  ~,
\label{eq:23}
\end{eqnarray}
where $\overline K_\omega$ is logarithmically divergent if $u_{\rm max}\to\infty$ while $\arctan\left( u_{\rm max}^2 /{\omega\tau_D}\right)\to\pi /2$ in $K_\omega$. The dependencies of $K_\omega /A$ versus $\omega\tau_D$ are shown in Fig. 2b [case (Aa)] for different aspect ratios and relaxation rates. $K_\omega /A$ is constant if $\omega <\nu$ and, for the wide intermediate spectral region $\nu <\omega <\tau_D^{-1}$, it is approximated by $0.6\ln (1/\sqrt{\omega\tau_D})$ (another possible fitting is $K_\omega\propto \omega^{-a}$ with $a\sim 0.4$.). If $\omega\tau_D\geq\widetilde u$, the  correlator decreases to $0.57p /\omega\tau_D$; we do not consider the region $\omega\tau_D\sim u_{\rm max}$ where $K_\omega\propto \omega^{-2}$. If $\omega\tau_D\leq\widetilde u$ or $\ell\leq\sqrt{\widetilde u}\ell_\omega$ ($\ell_\omega =2\sqrt{D/\omega}$ is the  diffusion length during times $\sim\omega^{-1}$), there is no dependency on aspect ratio and $K_\omega\propto (a\ell )^2\ln (\ell_\omega /\ell)$. By contrast, for the 1/f spectral region the integral contribution in Eq. (\ref{eq:23}) is weak [$\propto (\omega\tau_D )^2$], $K_\omega /A$ is not dependent on $\nu$, and the size dependence appears to be weaker, $K_\omega\approx 0.57Ap /\omega\tau_D\propto a\ell$. For the case (Ab), $\overline K_\omega /B$  is logarithmically divergent, if $u_{\rm max}^2\to\infty$, and it increases with $p$. Spectral dependencies vary slowly, from constant at $\omega\tau_D\leq 1$ to $\propto p\ln (u_{\rm max}/\sqrt{\omega\tau_D})$ at $u_{\rm max}\gg\sqrt{\omega\tau_D}\gg 1$ as it is plotted using semi-log scale in Fig. 2c for $\nu\tau_D =0.05$; these results remain weakly dependent on relaxation rate even for $\nu\tau_D\geq 1$. The size dependency of $\overline K_\omega$ is determined by the factor $Bp\ln (u_{\rm max}/\widetilde u)$, if $\omega\tau_D\leq\widetilde u$, and the $\ln$-factor should be replaced by $\ln (u_{\rm max}\ell_\omega /\ell )$ if $\omega\tau_D\geq\widetilde u$. Neglecting weak $\ln$-variations we obtain $\overline K_\omega\propto a\ell^3$, i. e. $\left\langle {\delta\Phi \delta\Phi}\right \rangle_\omega$ increases with TL's sizes faster than the area of the TL, $a\ell$.

\subsection{Noise excited at interfaces of wires }
We turn now to case (B) when random magnetization is localized at interfaces of superconducting wires. The magnetic field around the two-wire TL is governed by the quasistatic equation $\Delta_3 {\bf h}_{{\bf r}t}=0$. Below, we analyze this equation in the bipolar coordinate system, ${\bf r}= (x, \tau ,\sigma )$, with $-\infty <\tau <\infty$ and $0<\sigma <2\pi$ which are varied over the Y0Z plane as shown in Fig. 1c. If $a\gg r_w$, the outside-of-wire region corresponds $|\tau |<\ln (a/r_w )\equiv\tau_w$ and the line connecting wires is along $\sigma\to 0$. The boundary conditions at the left ($-$) and the right ($+$) wire's interfaces $S_\pm$, where $\tau\to\pm\tau_w$ and $0<\sigma <2\pi$, are written for tangential and normal components of the field as $h_{{\bf r}t}^{(\tau )}|_{S_\pm}=\widetilde h_{x\sigma t}^{(\pm )}$ and $h_{{\bf r}t}^{(\sigma )}|_{S_\pm}=0$, respectively. Here $\widetilde h_{x\sigma t}^{(\pm )}$ represents random magnetic fields at the interfaces and there are no fields normal to the superconducting wires. At $|{\bf r}|\to \infty$, we use the zero boundary condition. After the Fourier transform ( $k$ is now the wave vector along 0X), one obtains the quasi-static equation for the non-zero $\tau$-component of random magnetic field 
\begin{equation}
\left[ {\frac{{\partial ^2 }}{{\partial \tau ^2 }} + \frac{{\partial ^2 }}{{\partial \sigma ^2 }} - \frac{{(ka/2)^2 }}{{(\cosh \tau  - \cos \sigma )^2 }}} \right]h_{k\tau \sigma t}^{(\tau )}  = 0   \label{eq:24}
\end{equation}
with the periodic boundary condition over $\sigma$ and $h_{k, \pm \tau _w \sigma t}^{(\tau )}  =\widetilde h_{k\sigma t}^{(\pm )}$. The random flux $\delta\Phi_t$ is determined  through $h_{x\tau\sigma t}^{(\tau )}$ from  Eq. (\ref{eq:13}) in a form similar to Eq. (\ref{eq:16}):
\begin{equation}
\delta \Phi_t\!\! =\!\! -\!\!\int\limits_{-\tau_w}^{\tau_w}\!\! \frac{d\tau g_{\tau\tau}}{c}\!\!\left(\!\! \left. {h_{x\tau\sigma t}^{\bot}} \right|_{x=-\ell /2}^{x=\ell /2} +\!\!\int\limits_{-\ell /2}^{\ell /2}\!\! dx h_{x\tau\sigma t}^{\bot}\!\!\right)_{\sigma =0} ,  \label{eq:25}
\end{equation}
where $g_{\tau\tau}=a/(\cosh\tau +1)$ is the square root of the metric tensor. \cite{19}

In analogy to Eq. (\ref{eq:14}), the Fourier transform over $\sigma$ and $t$ for the tangential component of the field at interfaces of the $m$-th wire, $\widetilde h_{kn\omega}^{(m)}$ (with $n=0,\pm 1,\ldots$ and $m=\pm$ for a wire placed at around $\tau =\pm\tau_w$), is governed by the 2D diffusion equation 
\begin{equation}
\left[ i\omega +\nu_m -D_m\left( \frac{d^2}{dx^2} -\frac{n^2}{r_w^2}\right) \right]\widetilde h_{xn\omega}^{(m)}=\zeta_{xn\omega}^{(m)} ~,  
\label{eq:26}
\end{equation}
where $D_m$ and $\nu_m$ are the diffusion coefficient and relaxation rate of the $m$-th wire and $|x|<\ell /2$. The boundary conditions at the ends of the TL are $(\widetilde h_{xn\omega}^{(m)}/dx)_{x=\pm\ell /2}=0$. If the scale of correlations in $\left\langle {\zeta_{x\omega}^{(m)} \zeta_{x'\omega '}^{(m')}} \right\rangle$ is significantly greater than $r_w$, no contributions other than $\widetilde h_{xn=0\omega}^{(m)}\equiv\widetilde h_{x\omega}^{(m)}$ are essential. At the same time, we suppose that the scale of correlations is negligible in comparison to other scales ($a$, $\ell$, and the diffusion length $\ell_\omega$). Similarly to Eq. (\ref{eq:15}), we determine random sources in the right-hand side of Eq. (\ref{eq:26}) through the short-range correlation function given by 
\begin{equation}
\left\langle {\zeta_{x\omega}^{(m)} \zeta_{x'\omega '}^{(m')}} \right\rangle\!\! =\!\!\delta_{mm'}\frac{\delta (\omega +\omega ')}{(2\pi )^2}\!\!\left(\!\! w_m +\overline w_m\frac{\partial^2}{\partial x\partial x'}\!\!\right)\!\delta (x-x ')
~.   \label{eq:27}
\end{equation}
Here, $w_m$ and $\overline w_m$ stand for strengths of (a)- and (b)-type of noises along the $m$-th wire. \cite{20} The non-uniform solution of Eq. (\ref{eq:26}),
\begin{equation} 
\widetilde h_{x\omega }^{(m)} =-\int_{-\ell /2}^{\ell /2}{dx'}G_{xx'}^{(m)} {\zeta_{x'\omega }^{(m)}}/{D_m },  ~~~~~~~
\label{eq:28}
\end{equation}
is written through the Green's function with the $m$-dependent complex wave vector $\kappa_\omega =\sqrt{(i\omega +\nu_m )/D_m}$,
\begin{eqnarray} 
G_{xx'}^{(m)}  = [{\kappa_\omega  \sinh (\kappa_\omega\ell )}]^{-1} ~~~~~~~~~~~  \\
\times\left\{ {\begin{array}{*{20}c}
   {\cosh \kappa_\omega  \left( {\ell /2 + x} \right)\cosh \kappa_\omega  \left( {\ell /2- x'} \right),} & {x < x'}  \\
   {\cosh \kappa_\omega  \left( {\ell /2 + x'} \right)\cosh \kappa_\omega  \left( {\ell /2 - x} \right),} & {x > x'}  \\
\end{array}} \right. , \nonumber    \label{eq:29}
\end{eqnarray}
and a jump of the derivative of $G_{xx'}^{(m)}$ takes place at $x=x'$.  

After averaging over $\sigma$, Eq. (\ref{eq:24}) for a random field $h_{k\tau\omega}$ takes the form
\begin{equation}
\left( {\frac{{d^2 }}{{d\tau ^2 }} + V_\tau  } \right)h_{k\tau \omega }  = 0, ~~~ V_\tau = \frac{{(ka/2)^2 }}{{(\cosh \tau )^2  + 1/2}}
\label{eq:30}
\end{equation}
with the boundary conditions $h_{k\pm\tau_w\omega}=\widetilde h_{k\omega }^{(\pm )}$. Here, the potential $V_\tau$ is localized around zero, where $|\tau |<\tau_0\geq 1$ but $\tau_0\ll\tau_w$, and the region $(-\tau_0 ,\tau_0 )$ with $V_\tau\neq 0$ can be replaced by the boundary conditions 
\begin{equation}
\left. {\frac{{dh_{k\tau \omega } }}{{d\tau }}} \right|_{\tau =-\tau _0 }^{\tau =\tau _0 }\!\!\!\!  \approx\!\! \int\limits_{-\tau_0}^{\tau_0 }\!\! {d\tau } V_\tau  h_{k\tau = 0\omega}  \equiv\!\! v_k h_{k\tau  = 0\omega} ~
\label{eq:31}
\end{equation}
and $\left. h_{k\tau\omega}\right|_{\tau =-\tau_0}^{\tau =\tau_0}\!\!\approx 0$. Integration over $\tau$ in Eq. (\ref{eq:31}) gives the coefficient $v_k\approx 1.52\tau_w (ka/2)^2$. The linear-dependent on $\tau$ solution of the problem involving (\ref{eq:30}) and (\ref{eq:31}) should be substituted to Eq. (\ref{eq:25}) and after integrations over $\tau$ and $x$ one connects the  Fourier transform of the flux, $\delta\Phi_\omega$, and the random fields at the interfaces (\ref{eq:28}) as follows:
\begin{eqnarray} 
\delta\Phi_\omega =-\frac{1}{c}\!\!\int\!\!{dk} \!\left( {\left. {\ell e^{ - ikx} } \right|_{ - \ell /2}^{\ell /2}  + \!\!\int\limits_{-\ell /2}^{\ell /2}\!\!{dxe^{-ikx}} } \right)  \nonumber  \\  
\times \int_{-\tau_w }^{\tau _w }\!\! {d\tau }g_{\tau\tau} h_{k\tau\omega }   =  - \frac{{2a\ell}}{c}\!\!\int\!\!{dk} \alpha_k\left( {\widetilde h_{k\omega }^{(+)} +\widetilde  h_{k\omega }^{(-)} } \right) ,  \\
\alpha_k =\left(\frac{1}{k\ell}-i\right)\sin \left(\frac{{k\ell}}{2}\right)\frac{2+1.39v_k /\tau_w}{2+v_k} ~. ~~~~~ \nonumber   \label{eq:32}
\end{eqnarray}
Here, $\ell$-dependent factors in $\alpha_k$ appear due to integration along $x$ and $\widetilde h_{k\omega }^{(m)}$ should be written through the Fourier transform of the Green's function (29):
\begin{eqnarray} 
G_{kx}^{(m)}  = \int_{-\ell /2}^{\ell /2} \frac{dx'}{2\pi} e^{ikx'} G_{x'x}^{(m)}  = \frac{e^{ikx}  -F_{kx}^{(m)}}{2\pi (k^2  + \kappa_\omega ^2 )} ~, ~~~  \\   
 F_{kx}^{(m)}= \frac{ik \cosh \kappa_\omega (x+x_1 )}{\kappa_\omega\sinh \kappa_\omega\ell}\left. e^{ikx_1} \right|_{x_1 =-\ell /2}^{x_1 =\ell /2} ~,  \nonumber   
\label{eq:33}
\end{eqnarray}
which is a continuous function of $x$ (it is convenient to replace $x\leftrightarrow x'$ here and below).

Furthermore, the Fourier transform of the flux-flux correlator $\left\langle {\delta\Phi \delta\Phi}\right\rangle_\omega$ is obtained after substitution of Eqs. (\ref{eq:28}), (33) into (32) and averaging according to Eq. (\ref{eq:27}). The result takes the form   
\begin{equation} 
\left\langle {\delta\Phi \delta\Phi}\right\rangle_\omega\!\! =\!\! \sum\limits_m \!\!\int\limits_{-\ell /2}^{\ell /2}\!\! {\frac{dx}{\ell /2}} \left( A_m  \left| {\cal G}_{x\omega }^{(m)} \right|^2\!\! +B_m \left| {\overline{\cal G}_{x\omega}^{(m)} } \right|^2\right) ,    \label{eq:34}
\end{equation}
where the coefficients $A_m =w_m\ell^5 (a/2\pi D_m c)^2$ and $B_m =\overline w_m\ell^3 (a/\pi D_m c)^2$ are similar to the ones used in Sect. IIIA after replacements $W\to w_m\ell^2$ and $\overline W\to \overline w_m\ell^2$. The distribution of $\left\langle {\delta\Phi \delta\Phi}\right\rangle_\omega$ along the TL is determined by the dimensionless factors
\begin{equation}
\left| \begin{array}{*{20}c}
   {\cal G}_{x\omega }^{(m)}   \\
   \overline{\cal G}_{x\omega }^{(m)}  
\end{array} \right| = \int {dk} \alpha_k \left| {\begin{array}{*{20}c}
   {2G_{kx}^{(m)}/\ell }  \\  {\partial G_{kx}^{(m)}/\partial x}  
\end{array}} \right|    \label{eq:35}
\end{equation}
with ${\partial G_{kx}^{(m)}/\partial x}$ given by the derivative of Eq. (33). Below, we consider the case of identical wires, when $D$, $w$ and $\overline w$ are not dependent on $m=\pm$, so $\sum_m A_m$ or $\sum_m B_m$ should be replaced by the doubled  coefficients, ${\cal A}=2w\ell^3 (a\ell /Dc)^2$ or ${\cal B}=2\overline w\ell^3 (2a/Dc)^2$. After the integration over the complex $k$-plane in Eq. (\ref{eq:35}) we transform the distributions (\ref{eq:35}) into
\begin{equation} 
\left| {\begin{array}{*{20}c}{|{\cal G}_{x\omega}|^2}  \\ {|\overline{\cal G}_{x\omega } |^2}  
\end{array}} \right| \simeq 
\left| {\begin{array}{*{20}c}
   {\left| {\frac{2}{{(\kappa _\omega\ell )^2 }} + \frac{{0.69\sinh (\kappa _\omega  x)}}{{\tau _w \cosh (\kappa _\omega\ell /2)}}} \right|^2 }  \\
   {\frac{0.12}{\tau_w^2 } \left| {\frac{{\cosh (\kappa _\omega  x)}}{{\cosh (\kappa _\omega\ell /2)}}} \right|^2 }  \\
\end{array}} \right| ~  \label{eq:36}
\end{equation}
and these distributions are not dependent on the aspect ratio $p$ explicitly, see Appendix B for details. Thus, the final result $\left\langle {\delta\Phi \delta\Phi}\right\rangle_\omega ={\cal K}_\omega +\overline{\cal K}_\omega$ is obtained after straightforward integrations along the TL in the form 
\begin{eqnarray} 
{\cal K}_\omega ={\cal A}\left[\frac{1}{2(\omega \tau_D )^2}+ \frac{0.48}{\tau_w^2}\sqrt {\frac{2}{\omega \tau _D }} D_- (\sqrt {2\omega \tau _D })  \right]  ~,  \nonumber  \\
\overline{\cal K}_\omega ={\cal B}\frac{0.12}{\tau_w^2}\sqrt{\frac{2}{\omega \tau _D }}D_+ (\sqrt {2\omega \tau _D })  ~, ~~~~~~~~  
\label{eq:37}
\end{eqnarray}
where $D_{\pm}(s)=(\sinh s\pm\sin s)/(\cosh s+ \cos s)$ and where we restrict ourselves to the collisionless case $\nu /\omega\to 0$.
\begin{figure}
\begin{center}
\includegraphics[scale=0.2]{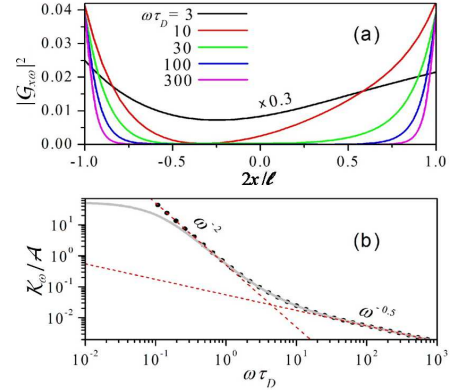}
\end{center}
\addvspace{-0.5 cm}
\caption{ (a) Density of the flux-flux correlator along a TL for case (Ba) versus $2x/\ell$ at different $\omega\tau_D$ (marked) and $\tau_m =3.5$. (b) Dimensionless correlator ${\cal K}_\omega /{\cal A}$ versus $\omega\tau_D$ at $\nu\tau_D =0$ (the dotted curve) and 0.1 (the gray solid curve). The asymptotes $\propto\omega^{-2}$ and $\propto\omega^{-0.5}$ are shown by dashed red lines for the low- and high-frequency regions, respectively. }
\end{figure} 

We discuss now spectral and size dependencies of the flux-flux correlator and its density distribution along the TL for cases (Ba) and (Bb), if $\nu\tau_D\to 0$ (saturation of noise for $\omega\leq\nu$ are shown in Figs. 3b and 4b). Distribution of the correlator density along the TL, $|{\cal G}_{x\omega }|^2$, versus the dimensionless coordinate $2x/\ell$ is plotted in Fig. 3a for case (Ba). If $\sqrt{\omega\tau_D}\gg 1$, the density of noise decreases exponentially around $|x|\sim 0$, but it is $\sim (0.69/\tau_m )^2$ in narrow regions near $|x|\sim \ell /2$. If $\sqrt{\omega\tau_D}\leq 1$, distribution becomes $x$-independent and noise increases $\propto (\omega\tau_D )^{-2}$. Due to this, spectral dependency takes form ${\cal K}_\omega  /{\cal A} \approx 0.5/(\omega \tau _D )^2$ in the low-frequency region while, in the high-frequency region, $\sqrt {\omega \tau _D }\gg 1$, noise decreases slower ${\cal K}_\omega  /{\cal A} \approx 0.67/\tau _m^2\sqrt{\omega \tau _D}$ because only narrow regions $|x|\sim \ell /2$ are essential. Such a behavior is in agreement with the numerical plots shown in Fig. 3b.  Size dependencies of these asymptotes are determined using the coefficients ${\cal A}$ and ${\cal B}$ as well as $\tau_D\propto\ell^2$ and $\tau_w =\ln (a/r_w)$. One obtains that ${\cal K}_\omega\propto (a/\ell )^2$ is only dependent on the aspect ratio for the low-frequency region while, for $\sqrt {\omega \tau _D }\gg 1$, the result ${\cal K}_\omega\propto a^2\ell /\tau_w^2$ increases with the sizes of the TL. 
\begin{figure}
\begin{center}
\includegraphics[scale=0.2]{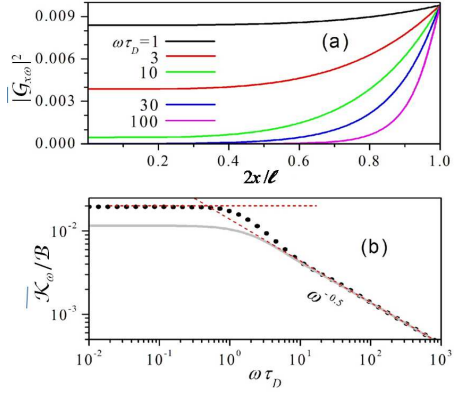}
\end{center}
\addvspace{-0.5 cm}
\caption{ (a) The same as in Fig. 3a for the case (Bb). (b) $\overline{\cal K}_\omega /{\cal B}$ versus $\omega\tau_D$ at $\nu\tau_D =0$ (the dotted curve) and 1 (the gray solid curve). High-frequency asymptote $\propto\omega^{-0.5}$ and low-frequency saturation are indicated by the dashed red lines for $\nu\tau_D =0$. }
\end{figure} 

Spectral and size dependencies for the case (Bb) are shown in Figs. 4a and 4b. At $\sqrt{\omega\tau_D}>1$, the distribution of the correlator along the TL, which is symmetric now ($\overline{\cal G}_{-x\omega}=\overline{\cal G}_{x\omega}$) due to the absence of an $x$-independent contribution in Eq. (\ref{eq:36}), decreases exponentially around $|x|\sim 0$. If $\sqrt{\omega\tau_D} <1$, this distribution becomes constant, $|\overline{\cal G}_{x\omega }|^2\approx 0.12/\tau_m^2$. As a result, $\overline{\cal K}_\omega /{\cal B}\approx 0.24/\tau_w^2$ in the low-frequency region, while, in the high-frequency region, $\sqrt {\omega \tau _D }\gg 1$, noise decreases as $\overline{\cal K}_\omega /{\cal B} \approx 0.17/\tau _w^2\sqrt{\omega \tau _D}$, cf Figs. 3b and 4b. Size dependencies of these asymptotic relations are determined with the use of the above coefficients. One obtains a value proportional to the square of the area dependency, $\overline{\cal K}_\omega\propto (a\ell )^2 /\tau_w^2$, in the low-frequency region and $\overline{\cal K}_\omega\propto a^2\ell$ appears for the high frequencies, the same as for case (Ba).

\section{Conclusions and outlook }
To summarize, we find that weak external noises should give a dominant contribution in long enough TLs. The flux-flux correlators considered in Sects. IIIA and IIIB show visible modifications of spectral dispersion in a wide spectral region around $\tau_D^{-1}$, and noise effect increases with the size of the TL, see asymptotics collected in Table I for short and long TLs at fixed $\omega$. The asymptotic behavior of $\left\langle {\delta\Phi\delta\Phi} \right\rangle_\omega$ in the low- and high-frequency regions is as follows:
\begin{eqnarray}
\left\langle {\delta \Phi \delta \Phi } \right\rangle _\omega   \approx A\left| \begin{array}{*{20}c}
   {1.4\ln \left( {\frac{1}{\sqrt{\omega \tau _D }}} \right)}  \\
   0.57p/\omega \tau _D  \\
\end{array} \right| + B0.72p\left| \begin{array}{*{20}c}
   {\ln \left({\frac{u_{\max }}{\widetilde u}} \right)}  \\
   {\ln \left( {\frac{{u_{\max } }}{\sqrt{\omega \tau _D }}} \right)}  \\
\end{array} \right| ~  \nonumber \\
 +{\cal A}\left| \begin{array}{*{20}c}
   0.5/(\omega \tau_D )^2   \\  
   \frac{0.67}{\tau _w^2 \sqrt{\omega \tau_D }} 
\end{array} \right| +{\cal B}\left| \begin{array}{*{20}c}
	0.24/{\tau_w^2 }  \\
   {\frac{0.17}{{\tau _w^2 \sqrt {\omega \tau _D } }}} 
\end{array} \right|, ~ \left| {\begin{array}{*{20}c}
   {\sqrt {\omega \tau _D }\ll 1}  \\
   {\sqrt {\omega \tau _D }\gg 1}  \\
\end{array}} \right| , ~~~  \label{eq:38}
\end{eqnarray} 
where the phenomenological parameters ($W$, $\overline W$, $w$, $\overline w$, $D$, and so on) should be determined from experimental data or microscopic models. These spectral and size dependencies enable us to separate different mechanisms of noise and opens the  way for an optimization and mitigation of relaxation effects on the coherent dynamics of superconducting devices used for quantum information processing. 
\begin{table}[t]
\centering
\begin{tabular}{|l| l| l|}
\hline 
$\left\langle {\delta \Phi \delta \Phi } \right\rangle _\omega$ & $~~~~ \ell\ll\ell_\omega$  &  $~~~~ \ell\gg \ell_\omega$  \\   \hline  
~~ (Aa) & $~~ \propto (a\ell )^2\ln (\ell_\omega /\ell )$ & $~~~~~~ \propto a\ell$ \\
~~ (Ab) & $\propto  a\ell^3\ln (u_{\rm max}a/2\ell )$ &  $\propto a\ell^3\ln (u_{\rm max}\ell_\omega /\ell )$ \\
~~ (Ba) & $~~~~ \propto (a/\ell )^2$ &  $\propto a^2 \ell /[\ln (a/r_w )]^2$ \\  
~~ (Bb) &  $\propto (a\ell )^2 /[\ln (a/r_w )]^2$  & $\propto a^2\ell /[\ln (a/r_w )]^2$ \\ \hline  
\end{tabular}
\caption{Size dependencies of $\left\langle {\delta \Phi \delta \Phi } \right\rangle _\omega$ for the cases under consideration. Here $\ell_\omega =2\sqrt {D/\omega}$ is the  diffusion length during times $\sim\omega^{-1}$. }
\end{table}

The results obtained are based on several assumptions, which are discussed below: \\
$\bullet~$ 
Phenomenological consideration of 1/f type noises in Sects. IIIA and IIIB are based on the diffusive Langevin equations (\ref{eq:14}) and (\ref{eq:26}) with the scalar coefficients $D$ and $\nu$ as well as with random sources which are determined through short-range correlators. External noises are considered without the effect of currents in the TL on random magnetization. A self-consistent microscopic study of these mechanisms and extension beyond the hydrodynamics approach require a special consideration. \\
$\bullet~$ 
In spite of we described the regime of classical noise in Sects. IIIA and IIIB, a similar consideration can be applied to the analysis of quantized random fields after replacement of the right-hand parts of Langevin equations (\ref{eq:14}) and (\ref{eq:26}) by quantized random sources. \\
$\bullet~$ 
Here, we restricted ourselves to the case of a TL shorted at $x=-\ell /2$; an open TL (with zero currents at $x=\pm\ell /2$) or the general case of a TL shunted by two-pole circuits at both ends should be checked separately using suitable boundary conditions for the telegraph equations. \\
$\bullet~$ 
We have considered here a simplest geometry of the direct two-wire TL (similar to the TL with fixed $\ell$ used in \cite{7} for characterization of spin defects at the interfaces). For a non-direct TL, the equation of motion in Sect IIA must be modified by taking into account any curvatures or angles along the TL. A coplanar TL requires additional numerical calculations, which can change the results quantitatively. \\
$\bullet~$ 
Equations of motion of the LC-shunted qubit in Appedix A are analyzed for frequencies lower than the characteristic frequency of the TL, $\omega_{LC}$. For frequencies comparable to $\omega_{LC}$, the problem of quantization cannot be based on the non-local equation of motion (\ref{eq:A1}), even when the classical regime of response becomes complicated and it should be analyzed starting with different assumptions. \cite{21} \\
$\bullet~$
The consideration performed here is irrelevant to a C-shunted qubit, \cite{22} where  the inductance contribution is negligible and the effect of flux noise on the dynamics of the qubit is reduced. However, TLs, which form quantum hardware based on these qubits, can be analyzed using the above results. \\
$\bullet~$ 
Finally, we have considered only low-frequency noises aroused from external magnetization described by the set of phenomenological coefficients, without an assumption of the thermodynamics equilibrium. A few additional mechanisms (noise due to dielectric losses, \cite{23} fluctuations of charge at the Josephson junction, \cite{24} interacting two-level defects, \cite{7,25} and fluctuations of current in the TL caused by thermal generation of electron-hole pairs \cite{26,27}) are possible for specific devices. The omitted mechanisms and the above-listed assumptions restrict the area of applicability of this paper but do not affect on the main result: the non-trivial spectral and size dependencies of the noise effect.

We turn now to a discussion of possibilities for experimental verification of the results obtained. Direct measurements of size-dependent flux-flux correlations are complicated but one can analyze  modifications of the spectral dependencies of a voltage-voltage correlator (\ref{eq:11}) or population relaxation time (\ref{eq:A6}) for long TLs in region $\omega\tau_D \sim 1$. Using a maximal value of diffusion coefficient $D\sim 10^{9}$ nm/s reported in \cite{12} and $\ell\sim 10~\mu$m, we estimate $\tau_D\sim 25$ ms so that frequency dispersion takes place in the spectral region $\leq$100 Hz. Such measurements allow one to separate different mechanisms of noise due to  the dependencies on the length of the TL (for the case of a narrow TL). It is also possible to mitigate relaxation effects, e.g. one can use a smaller TL area at fixed $\omega_{LC}$. The paper presents analytical results for a two-wire TL, which give explicate dependencies on the phenomenological parameters of noise and  open the way for the   qualitative verification of noise parameters in the TL via random voltage at the open end of a TL or via the population relaxation rate in a qubit shunted by a TL. Quantitative estimates for the structures used in current experiments, see \cite{12,13} and the references therein, or for a specially designed structure require additional numerical simulations for the verification of noise effects because field distributions around two-wire and coplanar TLs are different. 

In closing, the importance of noise effects in $\sim$100-qubit network with a length of TL $\sim$1 cm per qubit stems from a one meter total lenth of TLs which is necessary to maintain a coherent response of this cluster. It should be stressed that the fidelity of any multi-qubit system used in quantum information processing, when the inter-qubit connections and the qubit's control lines are based on different types of TL, is determined by noise levels in all elements of the device, both qubits and TLs. We believe that further study of noise-induced limitations will open the way to improving the parameters of the quantum hardware. 

\appendix   
\section{LC-shunted qubit}
A convenient way for verification of TL with noises is to study flux qubit formed by SQUID loop shunted by this TL, where the nonlinear current-voltage characteristics of Josephson junction provide a double-well potential \cite{28} with frequency level splitting in the GHz region. \cite{12,13,29} Equation of motion for such a qubit is determined from the Kirchhoff's requirement of current conservation,  $I_{J\omega}=I_{+,\omega}$, where $V_{+,\omega}$ in Eq. (\ref{eq:9}) should be replaced through flux at $\ell /2$ as $i\omega\Phi_{\omega}$. This edge condition is transformed into an equation of motion for a qubit by the use of the current through the Josephson junction $I_{Jt}=I_c \sin (2\pi \Phi_{Jt}/\Phi_0 )$, where $I_c$ is the critical current, $\Phi_0$ is the flux quantum, and $V_{Jt}+V_{x=0t}=0$ or $d\Phi_{Jt}/dt=-d\Phi_{x= 0t}/dt$ or $\Phi_{Jt}=-\Phi_{x=0t}$ (under a suitable initial condition). Note, that here we use a single-junction model with a variable effective critical current, $I_c$, instead of a   description of a SQUID with the total current through a loop which is controlled by external flux. In the $t$-domain one arrives to a non-local equation of motion $I_{Jt}=I_{+,t}$ which is written in the form
\begin{equation}
I_c\sin\frac{2\pi\Phi_t}{\Phi_0}=\int_{-\infty }^\infty {dt'}{\cal Y}_{t-t'}\frac{d\Phi_{t'}}{dt'}-\delta{\cal J}_{t}   \label{eq:A1}
\end{equation}
with the kernel ${\cal Y}_{t-t'}$ and random current $\delta{\cal J}_{t}$ determined by the Fourier transforms of Eqs. (\ref{eq:9}) and (\ref{eq:10}), respectively. 

Within the low-frequency limit, $\omega\ll\omega_{LC}$, the equation of motion (\ref{eq:A1}) takes the form:
\begin{equation}
\frac{C}{3}\frac{{d^2 \Phi _t }}{{dt^2 }} + \frac{{\Phi _t }}{L}+I_c \sin \frac{{2\pi \Phi _t }}{{\Phi _0 }} +\frac{\delta\Phi_t}{L}  =0   \label{eq:A2}
\end{equation}
and only third part of the TL's capacitance is involved here. The classic Hamiltonian, which results in this  equation of motion, can be written through charge-flux variables, $Q_t$ and $\Phi_t$, as follows
\begin{equation}
H_t  =\frac{3}{2}\frac{{Q_t^2 }}{C} + \frac{{\Phi _t^2 }}{{2L}}- \frac{{I_c \Phi _0 }}{{2\pi }}\cos \frac{{2\pi \Phi _t }}{{\Phi _0 }} +\frac{\Phi_t\delta\Phi_t}{L}  ~. \label{eq:A3}
\end{equation}
After the standard quantization procedure of $H_t$ which is based on the commutation relation $\left[ {\Phi ,\hat Q} \right] = i\hbar$, one arrives to the flux qubit Hamiltonian \cite{13,29}
\begin{equation}   
\hat H = 	\frac{3}{{2C}}\frac{{d^2 }}{{d\Phi ^2 }} + \frac{{\Phi ^2 }}{{2L}} - \frac{{I_c \Phi _0 }}{{2\pi }}\cos \frac{{2\pi \Phi }}{{\Phi _0 }}+\frac{\Phi\delta\Phi_t}{L} ,  \label{eq:A4}
\end{equation}
where an effective flux noise $\delta\Phi_t$ is dependent on the TL's parameters according to Eq. (\ref{eq:12}). Using the noiseless wave functions, which are determined by the eigenstate problem $\hat H_{\delta\Phi =0}|r\rangle =\varepsilon_r |r\rangle$ with $r=0,1,\ldots~$, we describe the qubit-environment interaction through the flux matrix elements $\Phi_{rr'}=\langle r|\Phi |r'\rangle$. Within the two-level approach ($r=0,1$) we use the $2\times 2$ Hamiltonian $\hat H_t =\hat\sigma_z\varepsilon_{10}/2+\hat \sigma_x\Phi_{10} \delta\Phi_t /L$ written through the Pauli matrices, $\hat{\bsigma}$, and the level splitting energy, $\varepsilon_{10}=\varepsilon_1 -\varepsilon_0$.  Here we consider the symmetric qubit, without any  tilt flux applied through the TL, and restrict ourselves to the classical noise regime, supposing that an effective temperature of noise $\delta\Phi_t$ is greater than $\varepsilon_{10}$.

In addition to a direct examination of $\left\langle{\delta\Phi\delta\Phi}\right\rangle_\omega$ via the voltage fluctuations (\ref{eq:11}), one can study relaxation of population in the flux qubit described by the Hamiltonian (\ref{eq:A4}). After averaging of the density-matrix equation over random noise, the balance equation for the populations of upper ($r=1$) and lower ($r=0$) levels takes the form \cite{30}
\begin{eqnarray} 
\frac{dn_{rt}}{dt} = ( - 1)^r\frac{\Phi _{10}^2}{L^2} \int_{ - \infty }^0 {d\Delta t} e^{\lambda \Delta t} \left\langle {\delta\Phi_{t +\Delta t} \delta\Phi_t }\right\rangle  \label{eq:A5}  \\
\times\left( {e^{i\omega_{10} \Delta t}  + e^{ - i\omega_{10} \Delta t} } \right)\left( {n_{0t+\Delta t}  - n_{1t+\Delta t} } \right)  ~,  \nonumber  
\end{eqnarray}
where $\omega_{10}=\varepsilon_{10}/\hbar$.
Within the Markov approximation $n_{rt+\Delta t}\approx n_{rt}$, the population re-distribution $\delta n_t =n_{1t}-n_{0t}$ is governed by the first-order equation $d\delta n_t /dt=-\delta n_t /T_1$ with the population relaxation rate
\begin{equation}
\frac{1}{T_1} \ =\frac{2\Phi _{10}^2}{L^2}\int_{ - \infty }^\infty  {d\Delta t} e^{i\omega_{10} \Delta t} \left\langle {\delta\Phi_{t +\Delta t}\delta\Phi_t }\right\rangle ~.  \label{eq:A6}
\end{equation}
Thus $T_1^{-1} =(2\Phi_{10}^2 /L^2)\left\langle {\delta\Phi\delta\Phi} \right\rangle_{\omega_{10}}$ and the relaxation time $T_1$ is connected with the correlator (\ref{eq:11}) according to the relation $\left\langle {\delta V\delta V} \right\rangle_\omega =(\omega L/\Phi_{10})^2 /2T_1$. Because the decoherence time $T_2$ is shorter than $T_1$ and characterization of the noise through $T_2$ is more complicated, we do not consider relaxation of the non-diagonal part of the density matrix.
\begin{figure}
\begin{center}
\includegraphics[scale=0.2]{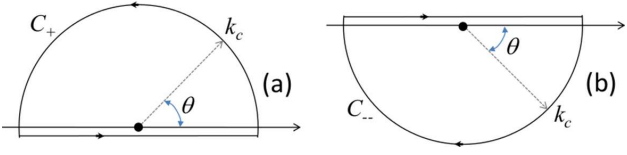}
\end{center}
\addvspace{-0.3 cm}
\caption{ Contours $C_+$ (a) and $C_-$ (b) used for integrations of the $\propto\exp (\pm k\ell /2)$ contributions in Eq. (\ref{eq:35}).  }
\end{figure} 

\section{Contour integration}
Here we calculate distributions (\ref{eq:35}) performing integrations over a complex $k$-plane with the use of the contours $C_{\pm}$ shown in Fig. 5. We separate the $\propto\exp (\pm k\ell /2)$ contributions to $\alpha_k$ in Eq. (32) and close integrals over $k$ in the upper/lower half-spaces, respectively. Under the substitution $G_{kx}$ given by Eq. (33) into ${\cal G}_{x\omega}$ one should take into account that there are no poles at $k\to\pm i\kappa_\omega$, i.e. $G_{\pm i\kappa_\omega x}<$const. The only pole appears because $\alpha_k\propto 1/k$ at $|k|\to 0$, so 
\begin{equation} 
\int\limits_{C_ \pm  } {\frac{{dk}}{i\ell}} e^{ \pm ik\ell /2} \alpha _k G_{kx}  =  \pm \frac{{2\pi }}{\ell^2}G_{k=0x}  = \pm (\kappa_\omega\ell )^{-2} 
\label{eq:B1}
\end{equation}
and ${\cal G}_{x\omega}$ is determined from the Cauchy's theorem ${\cal G}_{x\omega}+\Delta{\cal G}_{x\omega}=2/(\kappa_\omega\ell)^2$ where $\Delta{\cal G}_{x\omega}$ involves integrals over the upper and lower half-circles. Replacing $k$ by $k_c e^{i\theta}$ and taking into account that $k_c a\to\infty$ in the factor $\alpha_k$ we calculate $\Delta{\cal G}_{x\omega}$ as follows
\begin{eqnarray} 
\Delta{\cal G}_{x\omega }  = i\frac{1.38}{\tau _w}k_c \left( \int\limits_0^\pi  {d\theta e^{i\theta } e^{ik\ell} G_{kx} } \right. ~~~~ \\
\left. - \int\limits_0^{ - \pi } {d\theta e^{i\theta } e^{ - ik\ell} G_{kx} }  \right)_{k \to k_c e^{i\theta } }  \approx \frac{{0.69}}{{\tau _w }}\frac{{\sinh (\kappa _\omega  x)}}{{\cosh (\kappa _\omega\ell /2)}}  ~.
\nonumber  \label{eq:B2}
\end{eqnarray} 
Because of $k_c\to\infty$, only the regions $\theta\sim 0$ and $\sim\pm\pi$ are essential here. Collecting these integrals, one obtains the $|{\cal G}_{x\omega}|^2$ given by Eq. (\ref{eq:36}).

Distribution $\overline {\cal G}_{x\omega}$ is determined by Eq. (\ref{eq:35}) through the derivative
\begin{eqnarray} 
 \frac{\partial G_{kx}}{\partial x} =ik \frac{e^{ikx}-\widetilde F_{kx}}{2\pi (k^2 +\kappa_\omega ^2 )} ~~~~~~  \\   
\widetilde F_{kx}= \frac{\sinh \kappa_\omega (x+x_1 )}{\kappa_\omega\sinh \kappa_\omega\ell}\left. e^{ikx_1} \right|_{x_1 =-\ell /2}^{x_1 =\ell /2} ~,  \nonumber   
\label{eq:B3}
\end{eqnarray}
and there are no poles  at $\pm i\kappa_\omega$ (similar to the above case) as well as at $|k|\to 0$. As a result, integrations along $C_\pm$ give zero,
\begin{equation} 
\int\limits_{C_ \pm  }^{} {\frac{{dk}}{i\ell}} e^{ \pm ik\ell /2} \alpha _k \frac{{\partial G_{kx} }}{{\partial x}} = 0 ~. \label{eq:B4}
\end{equation}
On the other hand, this integral can be written through $\overline{\cal G}_{x\omega }$ and integrals over the upper and lower half-circles. Similarly to Eq. (\ref{eq:A2}) we obtain
\begin{eqnarray} 
\overline{\cal G}_{x\omega }  = i\frac{1.38}{\tau _w }k_c \left( \int\limits_0^\pi  {d\theta e^{i\theta } e^{ik\ell}\frac{\partial G_{kx}}{\partial x} } \right. ~~~~ \\
\left. - \int\limits_0^{ - \pi } {d\theta e^{i\theta } e^{ - ik\ell}\frac{\partial G_{kx}}{\partial x} } \right)_{k \to k_c e^{i\theta } }\!\!\approx \frac{1.38}{4\tau _w}\frac{{\cosh (\kappa _\omega  x)}}{{\cosh (\kappa _\omega\ell /2)}}
\nonumber  \label{eq:B5}
\end{eqnarray} 
and $|\overline{\cal G}_{x\omega }|^2$ is given by Eq. (\ref{eq:36}).


\begin{thebibliography}{42}
\bibitem{17} 
In contrast to Refs. 7 and 8, $\hat Y_\omega$ is Hermitian matrix because drop of voltages $V_{\pm ,~\omega}$ are determined between the same wires at $x=\ell /2$ and $x=-\ell /2$.
\bibitem{20} 
Dimensions of $w_m ,~\overline w_m$ and $W ,~\overline W$ in Eqs. (\ref{eq:15}) and (\ref{eq:27}) are  different by factor $\propto\ell^2$ and it leads to different size dependencies for the cases (A) and (B).
\end{thebibliography}
\end{document}